\documentclass[apj]{emulateapj}
\usepackage{graphicx}
\usepackage{amsmath,amssymb}
\usepackage{amsmath}

\def\mnras{MNRAS}
\def\apj{ApJ}
\def\aj{AJ}
\def\apjl{ApJL}

\let\oldAA\AA
\renewcommand{\AA}{\text{\normalfont\oldAA}}

\begin{document}

\title{The Contribution Of Outer HI Disks To The Merging Binary Black Hole Population}   

\author {Sukanya Chakrabarti  \altaffilmark{1}, 
Philip Chang  \altaffilmark{2}, 
Richard O'Shaughnessy \altaffilmark{1},
Alyson M. Brooks  \altaffilmark{3},
Sijing Shen \altaffilmark{4},
Jillian Bellovary  \altaffilmark{5},
Wojciech Gladysz \altaffilmark{6},
Chris Belczynski \altaffilmark{6}
}

\altaffiltext{1}
{School of Physics and Astronomy, Rochester Institute of Technology, 84 Lomb Memorial Drive, Rochester, NY 14623; chakrabarti@astro.rit.edu}
\altaffiltext{2}
{University of Wisconsin-Milwaukee}
\altaffiltext{3}
{Rutgers University, Department of Physics and Astronomy, 136 Frelinghuysen Rd, Piscataway, NJ 08854}
\altaffiltext{4}
{University of Oslo}
\altaffiltext{5}
{City University of New York}
\altaffiltext{6}
{Warsaw Univesity Observatory}

\begin{abstract}

We investigate the contribution of outer HI disks to the observable population of merging black hole binaries.  Like dwarf galaxies, the outer HI disks of spirals have low star formation rates and lower metallicities than the inner disks of spirals.  Since low-metallicity star formation can produce more detectable compact binaries than typical star formation, the environments in the outskirts of spiral galaxies may be conducive to producing a rich population of massive binary black holes.  We consider here both detailed controlled simulations of spirals and cosmological simulations, as well as the current range of observed values for metallicity and star formation in outer disks.   We find that outer HI disks contribute at least as much as dwarf galaxies do to the observed LIGO/Virgo detection rates.  Identifying the host galaxies of merging massive black holes should provide constraints on cosmological parameters and insights into the formation channels of binary mergers.  

\end{abstract}

\section{Introduction}

With the direct detection of gravitational waves from GW150914 (Abbott et al. 2016) began the new field of gravitational wave astronomy.  The signal arose from the merger of two massive  (with total mass of $60\,M_{\odot}$) stellar mass black holes that likely formed in a low-metallicity environment (Abbott et al. 2016; Belczynski et al. 2016).  Subsequently there have been two other detections -- a 22 $M_{\odot}$ merger (GW151226, Abbott et al. 2016) and a 50 $M_{\odot}$ merger (GW170104, Abbott et al. 2017),  as well as a likely candidate signal (LVT151012, Abbott et al 2016).  The masses of the black holes involved in these events strongly point to their formation in sub-solar metallicity environments, where inefficient winds allow massive stars to retain much more of their birth mass.  

The absolute luminosity distance provided by gravitational waves means that they can act as standard sirens (Schutz 1986), and the distances they provide do not depend on the astrophysical systematics of standard candles.  Space-based detectors like LISA are expected to provide accurate localization even without electromagnetic counterparts (Lang et al. 2011; Lang \& Hughes 2008), and future LIGO/Virgo runs are expected to produce some well-localized events (Chen \& Holz 2016).   The luminosity distance from gravitational waves, when combined with positional information on the host galaxy, can enable an independent measurement of cosmological parameters (Cutler \& Holz 2009), and provide insights into their formation channels.  With electromagnetic counterparts, as for neutron star mergers, one can obtain a wealth of information from multi-messenger studies (Abbott et al. 2017), and electromagnetic counterparts may be present for black hole mergers as well (de Mink \& King 2017). 

Binary black holes may form in several channels, including common envelope evolution (Belczynski et al. 2016; Kruckow et al. 2016), chemical homogeneous evolution (de Mink \& Mandel 2016, Marchant et al. 2016), and dynamical processes (Antonini \& Rasio 2016, Rodriguez et al. 2016, Chatterjee et al. 2017).  
Most of the theoretical work on the interpretation of the LIGO signal has focused on predicting merger rates per unit volume (Belczynski et al. 2008 and others), but there have been some recent works that have studied the host galaxies of binary black hole (BBH) mergers with binary merger population synthesis models (Lamberts et al. 2016; O'Shaughnessy et al. 2017; henceforth O17).   Lamberts et al. (2016) found that the signal is likely arising from either dwarf galaxies or massive galaxies at higher redshift.  O'Shaughnessy et al. (2017) pointed out that due to the low-metallicity star formation found in dwarf galaxies, binary black hole mergers would be abundantly produced in these systems. 

The environments of outer HI disks of massive galaxies are similar to dwarf galaxies -- their metallicities are lower than the inner disk (Ferguson et al. 1998; Kennicutt et al. 2003; Bresolin et al. 2009), and there is a low level of star formation in the outer disk (Thilker et al. 2005; Gil de Paz et al. 2007; Bigiel et al. 2010).    Outer HI disks have a flat radial profile (Wong \& Blitz 2002), and extend out to several times the optical radius.  They can contain a significant fraction of the atomic hydrogen mass in spiral galaxies (Bigiel et al. 2010).  Because HI disks are cold, they are sensitive to gravitational perturbations, and because of their extent, they present a large cross-section for interactions. Earlier works (Chakrabarti \& Blitz 2009; 2011; Chakrabarti et al. 2011 (henceforth C11); Chakrabarti 2013) have analyzed the morphology of the outer HI disks of the Milky Way and local spirals to characterize the galactic satellites (dwarf companions) and dark matter distribution in these systems.  We use some of the results from the simulations of M51 from C11 in this paper. 

Here, we compare the expected merger rate and detection rate of binary black holes formed in the outer HI disks of spirals with dwarf galaxies.   We summarize the star formation and metallicity values from observations and simulations in \S 2, and give our calculations for the merger rate and detection rate of BBHs formed in outer HI disks in \S 3.  We conclude in \S 4.

\section{Summary of Parameters from Observations and Simulations}

To calculate the detection rate of BBHs in outer HI disks and dwarf galaxies, we require knowledge of the galactic star formation rate ($\dot{M}_{\star}$) and metallicity ($Z$) in the outskirts.   We first discuss the values of these parameters as derived from observations.  We then summarize these parameters as derived from simulations.  

Although different works find that outer HI disks are sub-solar, there is a wide range of metallicity values that have been found from using various tracers and methods, and we briefly summarize the results here.  Ferguson et al. (1998) measured the gas-phase abundances of $O/H$ and $N/O$ in the outer regions of several spirals out to twice $R_{25}$ (the 25-th B-band magnitude isophote), and found that they were 10 - 15 \% of the solar value, and 20 - 20 \% of the solar value respectively.  Gil de Paz et al. (2007) found that the oxygen abundance is about 0.1 relative to solar in the outermost regions of the outer HI disk of M83.  Kennicutt et al. (2003) derived oxygen abundances of about 1/15 relative to solar in the outermost regions of M101.  Bresolin et al. (2009) find a discontinuity in the radial oxygen abundance in the outer HI disk of M83, and note that the metallicity is of order a third of the solar value (or higher depending on the diagnostic).  Bresolin et al. (2012) found the oxygen abundances in the outer HI disks of two spiral galaxies to be about 35 \% of the solar value.  Various tracers and methods yield somewhat different values for the metallicity in the outskirts (Kudritzki et al. 2012), and it is beyond the scope of this paper to discuss this in detail.  For the purposes of our work here, we note the range in observationally determined metallicities varies from 0.06 - 0.35 relative to solar in outer HI disks.  The relatively flat metallicity gradient seen in some outer disks (i.e. rather than exponentially declining as expected from simple closed box models) may be produced from non-circular flows that are efficient at redistributing metals in the disk (Yang \& Krumholz 2012; Petit et al. 2015).  The star formation rate in outer HI disks is observationally determined to be $\sim 0.05 - 0.1 M_{\odot}/\rm yr$ (Bigiel et al. 2010).

Both dwarf spheroidals and gas-rich dwarf irregulars have low metal content and lie on the mass-metallicity relation (Kirby et al. 2013; Lee et al. 2006).  In \S \ref{sec:detectrate}, we will adopt the functional form of the mean mass-metallicity relation from Panter et al. (2008), and the extension to dwarf galaxies by Lee et al. (2006), as well as the more recent work by Andrews \& Martini (2013).   Star formation rates in dwarf galaxies can vary from $\sim 10^{-3} M_{\odot}/\rm yr$ in HI-selected galaxies (Huang et al. 2012) to higher star formation rates ($\sim 10^{-2} - 10^{-1}~M_{\odot}/\rm yr$) for SDSS samples (Salim et al. 2005). 

O17 studied cosmological simulations of Milky-Way like galaxies that have been found to reproduce characteristics of typical spirals.  The details of the simulations are described in Brooks et al. (2007), Governato et al. (2009),  Bellovary et al. (2014), and Christensen et al. (2016).  Brooks et al. (2017) found that the outermost HI sizes in these simulations are matched to the rotational velocity.  Metals are produced in supernova explosions and a scheme for turbulent metal diffusion is included in these simulation (Shen et al. 2010).  The average metallicity in the outer HI disk of the h258 spiral galaxy simulation considered in O17 is 0.4 relative to solar, and metallicities in the outer disks of the other spiral galaxy simulations considered in O17 are similar.   For our dwarf galaxy calculation, we consider the h516 dwarf galaxy simulation from O17 as our fiducial case which has $\dot{M}_{\star,d} \sim 0.01~ \rm M_{\odot}/\rm yr$, $Z \sim 0.1 Z_{\odot}$, and stellar mass, $M_{\star,dwarf} = 2.5 \times 10^{8} M_{\odot}$.   The h603 dwarf in O17 has $\dot{M}_{\star,d} \sim 1~\rm M_{\odot}/\rm yr$, $Z \sim 0.4 Z_{\odot}$, and $M_{\star,dwarf} = 7.8 \times 10^{9} M_{\odot}$.  We also consider here the Eries cosmological simulation of a Milky Way like galaxy (Guedes et al. 2011); the metal production in that simulation has been described in detail in Shen et al. (2015).  For the Eries simulation, the mean metallicity in the outskirts is $Z = 0.06~Z_{\odot}$, and the star formation rate in the outskirts is $0.1 M_{\odot}/\rm yr$.

 C11's simulations are not cosmological, but by design match the detailed morphology of M51 (and also NGC 1512 that was also studied in that paper), and the gas disk in C11 extends out to large radii ($r \sim 50~\rm kpc$).  C11 used the Gadget-2 SPH code to carry out detailed controlled simulations of M51 and NGC 1512 and fit to the observed morphology of the outer HI disk to characterize the galactic satellites in these systems.  We consider C11's M51 simulation as our fiducial case here.  The average metallicity in the outskirts in C11's simulation is 0.1 relative to solar, and the average star formation rate in the outskirts is 0.1 $M_{\odot}/ \rm yr$, which is comparable to observed values (Thornley et al. 2006; Bigiel et al. 2010).

\section{Detection Rate of Binary Black Holes in Outer HI Disks and Dwarf Galaxies}
\label{sec:detectrate}

\begin{table*}
\centering
        \caption{\textbf{Merger Rate }}

          \begin{tabular}{@{}lcccc@{}}
          \hline

Source (HI Disks) 	                              &  	$Z/Z_{\odot}$ 	   &  $\lambda(Z)$      &    Merger Rate  &  Mass-Weighted Merger Rate\\	
\hline

Kennicutt et al. (2003)	                       &    0.0667                &    0.0004                     &    4.33        & 65          \\

Chakrabarti et al. (2011)                         &    0.1                      &    0.0004                  &       4.3            &  65       \\

Bresolin et al. (2012)                              &    0.35                     &    0.00027                    &     2.9           &  8.5      \\

O'Shaughnessy et al. (2017)                    &    0.4                      &     0.00023                     &  2.5              &   6.2     \\

Shen et al. (2015)                                  &   0.06                     & 0.0004                    &   4.3             & 65       \\

\hline
Source (Dwarf Galaxies)           \\ 
\hline

h516 dwarf from O17                           &   0.1   (Panter et al. 2008)    &  0.0004                      &   0.9   &   3.9 \\
h516 dwarf from O17                           &   0.1   (Andrews \& Martini 2013)     & 0.0004          &    0.9     &  0.8 \\
h516 dwarf from O17                           &   0.1   (Lee et al. 2006)                    & 0.0004          &    0.9     &  1.7 \\
h603 dwarf from O17                            & 0.4 (Panter et al. 2008)        & 0.0002                         &  1.5    &  6.5 \\
h603 dwarf from O17                           & 0.4 (Andrews \& Martini 2013)     & 0.0002               & 1.5         &  1.4 \\
h603 dwarf from O17                           & 0.4 (Lee et al. 2006)                    & 0.0002               & 1.5         &  2.8 \\

\hline

\end{tabular} 

\small{The first column gives the reference that we have used for $Z_{\rm HI}$ or $Z_{\rm dwarf}$; second column gives the reference metallicity ($Z_{\rm HI}$ or $Z_{\rm dwarf}$), as well as the mass-metallicity relation that we adopt for dwarf galaxies, third column gives the metallicity-dependent mass efficiency, in units of [$1/M_{\odot}$], the fourth column gives the merger rate, and the fifth column gives the mass-weighted merger rate, both of these quantities are given in units of [$\rm Gpc^{-3}~\rm yr^{-1}$].}

\end{table*} 

To determine the merger rate of binary black holes for a given star formation rate ($\dot{M}_{\star}$) and metallicity ($Z$) in outer HI disks, we carry out a similar calculation as in O17.  We begin by considering M51 as a prototypical spiral with an extended HI disk, and we adopt parameters from the simulation of M51 performed by C11 as our fiducial numbers.   The merger rate of BBHs in outer HI disks is denoted $\dot{n}_{\rm HI}$ and is given by:

\begin{equation}
\dot{n}_{\rm HI} =\int d M_{HI} \phi(M_{HI}) \left(\frac{M_{HI}}{M_{HI, M51}}\right) \dot{M}_{\star,M51}~\lambda(Z) P (t < t_{\rm H}),  \\ 
\end{equation}

where $\lambda(Z)$ is the mass efficiency, or the amount of mass in stars that forms black holes, and depends on the metallicity $Z$, $P (t < t_{H})$ is the fraction of binaries that are merging now and incorporates a delay time distribution as discussed in prior papers (O17), $M_{HI}$ is the HI mass, $M_{HI,\rm M51}$ is the HI mass of M51, $\phi(M_{HI}$) is the HI mass function, adopted from Martin et al. (2010), and $\dot{M}_{\star,M51}$ is the star formation rate of M51.  The limits of integration are $10^{8} - 10^{11} M_{\odot}$ and correspond to the range of HI masses of spirals in surveys of the low redshift universe (Walter et al. 2008; Catinella et al. 2010).  We use the value of the HI mass of M51 given in Leroy et al. (2008), $M_{HI, M51} = 3.16 \times 10^{9} M_{\odot}$.   To determine the total merger rate from the population of HI disks, we integrate over HI masses, using the HI mass function, $\phi(M_{HI})$, from Martin et al. (2010) that gives the HI mass function for galaxies in the local universe.   From C11, we adopt $\dot{M}_{\star,M51} = 0.1 M_{\odot}/\rm yr$ and $Z = Z_{\rm M51} = 0.1~Z_{\odot}$ for the star formation rate and metallicity of M51 respectively.  Here, we assume that the star formation rate of the outer HI disk is proportional to its HI mass.  Bigiel et al. (2010) found that the star formation in the outskirts ($r > r_{25}$) is linearly proportional to the HI surface density (and thus to the HI mass) for a constant depletion time.   This is not the case in the molecule dominated region (Bigiel et al. 2008; Popping et al. 2014) for galaxies as a whole.

\begin{figure}[h]
\begin{center}
\includegraphics[scale=0.45]{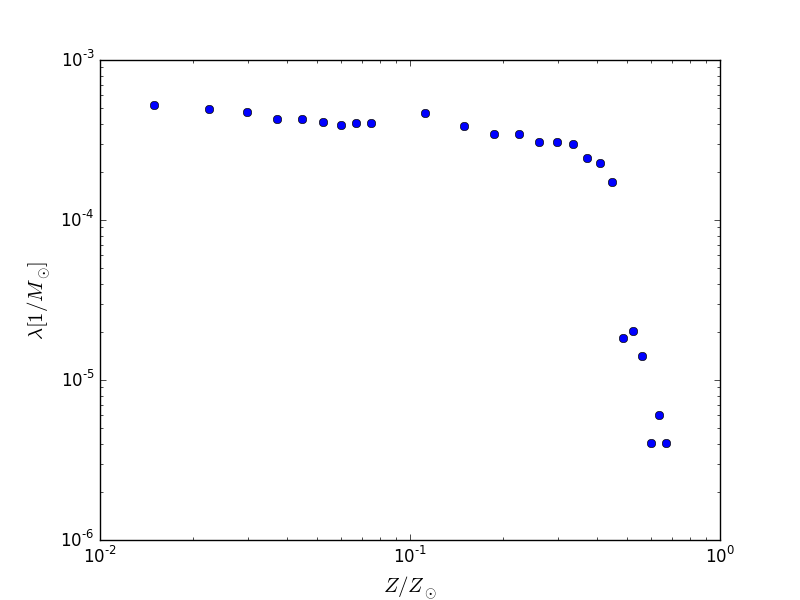}
\caption{The mass efficiency $\lambda$, or fraction of mass in stars that forms black holes, as a function of metallicity relative to solar, adopted from Belczynski et al. (2016), for binary black hole masses $> 40 M_{\odot}$. \label{f:lambda}}
\end{center}
\end{figure}

We adopt values for $\lambda$, the mass efficiency, from Belczynski et al.'s (2016) binary merger population synthesis model for a range of metallicities.  Figure \ref{f:lambda} depicts the mass efficiency as a function of metallicity.  For comparison to the GW15014 event and the other massive BBH mergers detected, here we plot the mass efficiency for BBH masses greater than 40 $M_{\odot}$.  The merging fraction, $P(t < t_{H})$, is adopted from O17 and can be expressed as:
\begin{equation}
P(t < t_{H}) = \int_{t_H}^{t} dt \frac{dP}{dt} (t-t_{b}) \sim 10^{-3} \; .
\end{equation}
where $t_{b}$ is the birth time and $dP/dt (t-t_{b})$ is the delay time distribution, which is approximately proportional to $1/t$ (O17), i.e., a significant fraction of mergers occurs shortly after their births as stars.  Here, we assume for simplicity that the star formation rate is constant, and that the merger time scale is short compared to the timescale over which the star formation rate varies.  O'Shaughnessy et al. (2012) have shown that incorporating a time-dependent cosmic star formation history and integrating over the delay time distribution leads a factor of two difference relative to our approach here.  Our approximation will break down if the star formation history varies dramatically, i.e. if the star formation rate evolves faster than the merger timescale, or if the delay time distribution is substantially different than O17's and other related works (Belczynzski et al. 2016).  We have tabulated in Table 1 the values that we adopt for the mass efficiency at the reference metallicities, which are obtained from a linear interpolation of the data from Belczynski et al. (2016).  The merging fraction, $P(t < t_{H})$, is taken to be $10^{-3}$ (O17).

 HI disks do not have the same metallicity as the galaxy itself and are sub-solar.  If we adopt $Z = Z_{\rm M51} = 0.1 Z_{\odot}$ from C11 to be representative of the population of HI disks, we would get a merger rate of $4.3 ~ \rm Gpc^{-3} ~ yr^{-1}$.  The range in observationally determined metallicities varies from 0.06 - 0.35 relative to solar in outer HI disks, and corrresponds to a merger rate of BBHs in HI disks that varies from $4.5 - 2.5~ \rm Gpc^{-3} ~ yr^{-1}$ respectively.  The range of metallicities from observations (the maximum and minimum values) along with the references, the cosmological simulations studied by O17 and Shen et al. (2015), and the controlled simulations performed by C11 are summarized in Table 1, along with the resultant merger rate.

To compare the merger rate that we have obtained for outer HI disks to the work done earlier by O17 for dwarf galaxies, we repeat the above calculation using the parameters from cosmological simulations of dwarf galaxies adopted by O17.   We use the mass-metallicity relationship to express $Z(M_{\star})$ for dwarf galaxies, and integrate over stellar masses to obtain the merger rate of BBHs formed in dwarf galaxies, $\dot{n}_{\rm d}$:

\begin{equation}
\dot{n}_{\rm d} = \int \rm d M_{\star} \phi(M_{\star}) \left(\frac{M_{\star}}{M_{\star, dwarf}}\right) \dot{M}_{\star,d}~\lambda(Z) P (t < t_{\rm H}) \\   \; .
\end{equation}

Here, $\dot{M}_{\star,d}$ is the star formation rate of a prototypical dwarf galaxy (adopted from the cosmological simulations of dwarf galaxies reported in O17), $\phi(M_{\star}$) is the galaxy stellar mass function, and we use the best-fit Schecter parameters from Fontana et al. (2006), evaluated locally ($z = 0$).   The limits of integration correspond to the range of stellar masses of dwarf galaxies in Lee et al.'s (2006) extension of the mass metallicity relation to dwarf galaxies, i.e. $10^{6} - 10^{9.3} M_{\odot}$.   
Evaluating this integral then gives merger rates of $\dot{n}_{\rm d} \sim 1 ~\rm Gpc^{-3} ~ yr^{-1} $ using the various mass-metallicity relations and are listed in Table 1.   As Lee et al.'s (2006) work extends the mass-metallicity relation specifically to the low-mass end, it is the most relevant mass-metallicity relation for use for dwarf galaxies.  Mannucci et al. (2010) found there is a fundamental relation between metallicity, star formation rate, and mass.  However, in the low metallicity regime, this relation is not well calibrated.  Here, we implicitly assume that the star formation rate is proportional to the stellar mass; more detailed studies should be done in the future that do not use this assumption.

\begin{figure}[h]
\begin{center}
\includegraphics[scale=0.48]{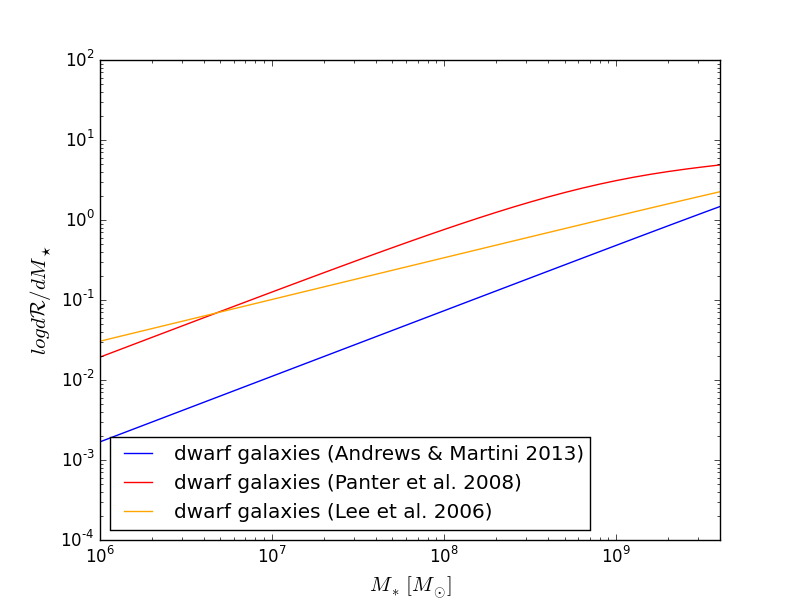}
\includegraphics[scale=0.48]{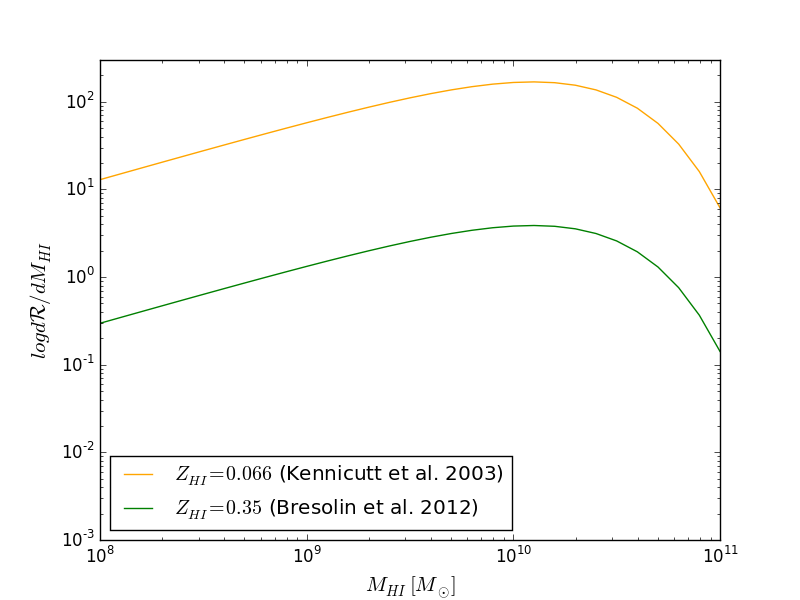}
\caption{(a)Top: Mass-weighted merger rate of BBHs formed in dwarf galaxies.  We depict the mass-weighted merger rate for dwarfs using the Panter et al. (2008) mean mass-metallicity relation, as well as the Andrews \& Martini (2013), and the Lee et al. (2006) mass-metallicity relations.  (b) Bottom: For outer HI disks, we have shown the mass-weighted merger rate for the observed range of metallicities in outer HI disks.   \label{f:detect}}
\end{center}
\end{figure}

Dominik et al. (2012) have described how metallicity influences both the merger rate [Eq. (2)] and also
   the characteristic mass of the merging BH-BH population.  
Since the detection volume scales roughly as $m_{chirp}^{15/6}$, where $m_{chirp}$ is the characteristic (chirp) mass of the merging
BH-BH population, the detection volume scales with metallicity as $V(Z) \propto Z_{\odot}/Z$ for any noise power spectrum. 
To characterize
   the relative impact of different populations on the GW detection rate, without making concrete assumptions
   about the detector network involved (e.g., without adopting a specific detector noise power spectrum and
   threshold), O17 simply rescaled the merger rate by this factor, producing what we denote a "mass-weighted
   merger rate":  

\begin{equation}
\label{eq:mergerrate}
\mathcal{R} = \int \rm d M_{i} \phi(M_{i}) \left(\frac{M_{i}}{M_{ref}}\right) \dot{M}_{\star, i}~ \lambda(Z) \left(\frac{Z_{\odot}}{Z}\right) P (t < t_{\rm H}) \\ ,
\end{equation}

where $i$ refers to HI or stars, and "ref" is the appropriate fiducial reference.  This quantity is proportional to the expected gravitational wave detection rate,
   with a proportionality constant   $V_{\rm ref}$, which  is the (average) volume to which a  network is sensitive
   to the population of binary black holes produced from solar-metallicity star formation. The mass-weighted merger rate is plotted in Figure \ref{f:detect} for both HI disks and dwarf galaxies.    The values of the mass-weighted merger rate are also tabulated in Table 1 for both outer HI disks and dwarf galaxies.  The average value from outer HI disks of 42 $\rm Gpc^{-3}~yr^{-1}$ is within the range of current LIGO's estimates of $103^{+110}_{-63}$ $\rm Gpc^{-3}~yr^{-1}$ (Abbott et al. 2017).  Normalizing by $V_{\rm ref} \sim 0.1~\rm Gpc^{3}$ (Abbott et al. 2016), we would expect a detection rate of $\sim$ 0.6 - 6 events per year from BBHs in outer HI disks so far, which is consistent with the number of events that have been observed so far.



\section{Discussion \& Conclusion}

Our calculation is approximate, and as Elbert et al. (2017) have pointed out, one cannot at present carry out an ab initio calculation of the BBH merger rate.  Thus, various works have adopted approximations.  Lamberts et al. (2016) used binary merger population synthesis models along with a mass-metallicity relation from a cosmological simulation, and abundance matching to determine the galaxy halo mass for each stellar mass, which may be problematic on the low mass end (Miller et al. 2014).   The total rate they obtain is $150~\rm Gpc^{-3} yr^{-1}$, and in the mass interval of dwarf galaxies is larger than what we find by a factor of several.  O17 used binary merger population synthesis models in conjunction with cosmological simulations, but did not calculate the merger rate in physical units, and did not study outer HI disks.  Elbert et al. (2017) used single stellar evolution models and parameterized the effects of binary formation with two parameters.  Their rates for dwarf galaxies for prompt massive BBH mergers are comparable to what we find here.  The environments of dwarf galaxies are similar to outer HI disks, but have not been studied in prior papers on BBH merger rates.

In summary, for the observed range of metallicities in outer HI disks,
  we estimate that BBHs formed in outer HI disks have a mass-weighted merger rate of 6 - 65 $~\rm Gpc^{-3} yr^{-1}$, and considering the range of adopted parameters from simulations gives a comparable number.  The average mass-weighted merger rate (averaged over parameters adopted from observations and simulations) of BBHs formed in outer HI disks is 41$~\rm Gpc^{-3} yr^{-1}$, and the average mass-weighted merger rate for the dwarf galaxy simulations considered here is $3~\rm Gpc^{-3} yr^{-1}$.  Thus, we expect that BBHs formed in outer HI
  disks may be a significant contributor to the observed  LIGO signal.
 
If there are electromagnetic counterparts for BBHs, it will be easier to localize a merger in a large host galaxy rather than a small, faint dwarf galaxy.  If the signal can be localized, then the absolute luminosity distance from gravitational waves will provide independent constraints on the Hubble parameter, as we have already seen for neutron star mergers (Abbott et al. 2017).

\bigskip
\bigskip

\acknowledgments

We thank the anonymous referee for many helpful comments on our paper.  SC gratefully acknowledges NSF grant no. 1517488 and NASA grant no. 16-ATP16-0197.  P.C. gratefully acknowledges support from the NASA ATP program through NASA grant NNX13AH43G, and NSF grant AST-1255469. 

\end{document}